# Observation of two-dimensional superlattice solitons

M. Heinrich,[1] Y. V. Kartashov,[2] L. P. R. Ramirez,[1] A. Szameit,[3] F. Dreisow,[1] R. Keil,[1] S. Nolte,[1] A. Tünnermann,[1] V. A. Vysloukh,[2] and L. Torner[2]

[1]*Institute of Applied Physics, Friedrich-Schiller-University Jena, Max-Wien-Platz 1, 07743 Jena, Germany*

[2]*ICFO-Institut de Ciencies Fotoniques, and Universitat Politecnica de Catalunya, Mediterranean Technology Park, 08860 Castelldefels (Barcelona), Spain*

[3]*Physics Department and Solid State Institute, Technion, 32000 Haifa, Israel*

We observe experimentally two-dimensional solitons in superlattices comprising alternating deep and shallow waveguides fabricated via the femtosecond laser direct writing technique. We find that the symmetry of linear diffraction patterns as well as soliton shapes and threshold powers largely differ for excitations centered on deep and shallow sites. Thus, bulk and surface solitons centered on deep waveguides require much lower powers than their counterparts on shallow sites.

Since the prediction of discrete solitons in waveguide arrays [1] and their experimental demonstration [2,3] such states were investigated in a number of settings [4-7]. Particular attention has been paid to the precise tuning of soliton properties. Periodic systems with complicated transverse shapes such as photonic superlattices (SLs) open new opportunities for soliton control [8]. Due to their binary unit cell, such structures provide the possibility to engineer a mini-gap within the first propagation band [9]. A variety of linear [10-14] and nonlinear phenomena, including formation of gap solitons [9,15], solitons in Bragg gratings [16], and defect gap solitons [17], were demonstrated in SLs. However, experimental investigations of SL solitons were limited to one-dimensional (1D) settings [18,19], so that the features of two-dimensional (2D) entities remain unseen to date. In this Letter, we report on experimental observation of solitons in 2D binary SLs. We show how soliton formation is affected by the choice of the excited sublattices as well as by the presence of surfaces.

To gain insight into the dynamics of soliton formation, we describe the propagation of light with the nonlinear Schrödinger equation for the dimensionless field amplitude $q$ assuming CW illumination:

$$i\frac{\partial q}{\partial \xi} = -\frac{1}{2}\left(\frac{\partial^2 q}{\partial \eta^2} + \frac{\partial^2 q}{\partial \zeta^2}\right) - |q|^2 q - R(\eta,\zeta)q. \qquad (1)$$

Here $\eta, \zeta$ and $\xi$ are the transverse and longitudinal coordinates normalized to characteristic transverse scale and diffraction length, respectively. The refractive index profile is given by $R(\eta,\zeta) = p_1 \sum_{n,m=-N}^{N} G(nd,md) + p_2 \sum_{n,m=-(N-1)}^{N} G(nd-d/2, md-d/2)$, where $p_1$ and $p_2$ represent the depths of centered and shifted sublattices, respectively. The separation between sites in each sublattice is designated $d$, while $G(\eta_k,\zeta_k) = \exp[-(\eta-\eta_k)^2/w_\eta^2 - (\zeta-\zeta_k)^2/w_\zeta^2]$ describes the elliptical shape of the individual waveguides with widths $(w_\eta, w_\zeta)$. Among the conserved quantities of Eq. (1) is the energy flow $U = \int\int_{-\infty}^{\infty} |q|^2 \, d\eta d\zeta$.

In accordance with the experiments we set $N = 3$ (i.e. lattices with 85 waveguides), $d = 6.4$ (64 μm) and $w_\eta = 1.1$, $w_\zeta = 0.3$ (11 μm × 3 μm). In the following, the case $p_1 > p_2$ is referred to as $D$-lattice, because central, corner, and edge sites belong to "deeper" sublattice, i.e. the lattice with higher refractive index. Analogously, the case $p_1 < p_2$ is termed $S$-lattice, with above mentioned sites then belonging to the "shallower" sublattice. Further we set $p_1 = 3.00$, $p_2 = 2.86$ for $D$-lattice and $p_1 = 2.86$, $p_2 = 3.00$ for $S$-lattice. The value $p_{1,2} \sim 3$ is equivalent to actual refractive index modulation depth of $\delta n \sim 3.3 \times 10^{-4}$.

We search for stationary solutions of Eq. (1) residing in the center, edge, or in the corner in the form $q = w(\eta,\zeta)\exp(ib\xi)$, where $b$ is the propagation constant. The dependence $U(b)$ for such solutions is non-monotonic in both $S$- and $D$-lattices. The branches where $dU/db \leq 0$ are unstable, while $dU/db > 0$ corresponds to stable solitons. 2D SL solitons exist above a cutoff propagation constant $b_{co}$ and corresponding threshold power $U_{th}$. In contrast to the findings in truncated 1D SLs [19], solitons in 2D $D$-lattices require considerably lower threshold powers for their existence [Fig. 1(a)]. Thus, for the parameters stated above the soliton residing in the central site of the $D$-lattice has a threshold of $U_{th} \approx 0.550$, while its $S$-lattice counterpart exists only above $U_{th} \approx 0.929$. This difference is remarkable, taking into account the small detuning between sublattices of only $|p_2 - p_1|/p_2 \sim 5\%$. The difference in threshold powers grows rapidly with increase of $|p_2 - p_1|$. In both lattice types corner solitons feature the lowest, while center solitons feature highest threshold [Fig. 1(b)]. Notice that due to the finite number of waveguides in the lattice, $b_{co}$ is slightly lower in $D$-lattices.

Representative profiles of 2D SL solitons are shown in Fig. 2. They expand across the lattice and undergo pronounced shape oscillations as $b \to b_{\text{co}}$. However, solitons in the $D$-lattice expand dramatically at $b_{\text{co}}$, covering almost the entire lattice [Figs. 2(a-c)], while their $S$-lattice counterparts extend over only a few neighboring sites when $b \to b_{\text{co}}$ [Figs. 2(d-f)]. This difference in soliton shapes becomes more pronounced with growing detuning $|p_2 - p_1|$ between sublattices. Yet, in both $S$- and $D$-lattice an increase of the propagation constant eventually results in the contraction of solitons into the initially excited lattice site as shown for the $S$-lattice in Figs. 2(g-k).

Our experiments were conducted in SLs with above discussed parameters fabricated via femtosecond-laser direct writing [20] in a fused silica sample with a length of 105 mm. Specific fabrication parameters are discussed in [21]. Waveguides were excited with a Ti:Sapphire laser system delivering 200 fs pulses at a wavelength of 800 nm with a repetition rate of 1 kHz. The resulting patterns at the output facet were imaged onto a CCD camera.

Figures 3 and 4 show the output intensity distributions at specific powers for excitation of the central, edge and corner sites of $D$- and $S$-lattice, respectively. For comparison, the first rows of both figures show the simulated patterns in the linear regime (vanishing excitation power), while the respective experimental linear patterns (peak power 200 kW) are depicted in the second row. In the third and fourth rows different stages of localization for peak powers of 1 MW and 2 MW are shown.

Notice that while near-threshold stationary solutions generally feature a lower degree of localization in the $D$-lattice, experimental observations suggest that upon dynamical excitation in the nonlinear regime the light spreads more widely in the $S$-lattice at comparable power levels. Figures 3 and 4 illustrate that experimental localization in $D$-lattices is achieved at substantially lower input peak powers than in $S$-lattices. While an increasing input power results in the monotonic contraction of the observed output pattern in $D$-lattices, an intermediate spreading occurs in $S$-lattices when due to nonlinear contributions the effective index of excited site reaches the value of its neighboring guides in the high-index sublattice. A similar effect was observed for the excitation of surface solitons on negative defects [22]. Since all experiments were conducted with pulsed light, the nonlinear index matching may occur on the pulse slopes even for peak powers well above the threshold [23]. This results in a more pronounced background and considerably lowers soliton excitation efficiency in $S$-lattices.

In conclusion, we observed experimentally the formation of solitons residing in the center, edge and corner sites of binary SLs for both $S$- and $D$-type configurations. We showed numerically that even a small refractive index offset between the sublattices has a strong influence on the respective power thresholds. Furthermore, both linear diffraction patterns and soliton profiles may differ substantially in lattices of different types.

# References with titles

1. D. Christodoulides and R. Joseph, "Discrete self-focusing in nonlinear arrays of coupled waveguides," Opt. Lett. **13**, 794 (1988).
2. H. Eisenberg, Y. Silberberg, R. Morandotti, A. Boyd and J. Aitchison, "Discrete spatial optical solitons in waveguide arrays," Phys. Rev. Lett. **81**, 3383 (1998).
3. J. Fleischer, M. Segev, N. Efremidis and D. Christodoulides, "Observation of two-dimensional discrete solitons in optically induced nonlinear photonic lattices," Nature **422**, 147 (2003).
4. F. Lederer, G.I. Stegeman, D.N. Christodoulides, G. Assanto, M. Segev and Y. Silberberg, "Discrete Solitons in Optics," Phys. Rep. **463**, 1 (2008).
5. Y. V. Kartashov, V. A. Vysloukh, and L. Torner, "Soliton shape and mobility control in optical lattices," Prog. Opt. **52**, 63 (2009).
6. X. Wang, A. Bezryadina, Z. Chen, K.G. Makris, D.N. Christodoulides, and G.I. Stegeman, "Observation of Two-Dimensional Surface Solitons," Phys. Rev. Lett. **98**, 123903 (2007)
7. A. Szameit, Y.V. Kartashov, F. Dreisow, T. Pertsch, S. Nolte, A. Tünnermann, and L. Torner, "Observation of Two-Dimensional Surface Solitons in Asymmetric Waveguide Arrays," Phys. Rev. Lett. **98**, 173903 (2007)
8. Y. S. Kivshar and N. Flytzanis, "Gap solitons in diatomic lattices," Phys. Rev. A **46**, 7972 (1992).
9. A. A. Sukhorukov and Y. S. Kivshar, "Discrete gap solitons in modulated waveguide arrays," Opt. Lett. **27**, 2112 (2002).
10. H. Ohno, E. E. Mendez, J. A. Brum, J. M. Hong, F. Agulló-Rueda, L. L. Chang, and L. Esaki, "Observation of Tamm states in superlattices," Phys. Rev. Lett. **64**, 2555 (1990).
11. J. Klos, "Tamm and Shockley states in two superlattices coupled by a perturbed interface," Phys. Stat. Sol. B **242**, 1399 (2005).
12. M. Ghulinyan, C. J. Oton, Z. Gaburro, L. Pavesi, C. Toninelli, and D. S. Wiersma, "Zener Tunneling of Light Waves in an Optical Superlattice," Phys. Rev. Lett. **94**, 127401 (2005).
13. N. Malkova, I. Hromada, X. Wang, G. Bryant, and Z. Chen, "Observation of optical Shockley-like surface states in photonic superlattices," Opt. Lett. **34**, 1633 (2009).

14. F. Dreisow, A. Szameit, M. Heinrich, T. Pertsch, S. Nolte, A. Tünnermann, and S. Longhi, "Bloch-Zener Oscillations in Binary Superlattices," Phys. Rev. Lett. **102**, 076802 (2009).
15. Y. J. He, W. H. Chen, H. Z. Wang, and B. A. Malomed, "Surface superlattice gap solitons," Opt. Lett. **32**, 1390 (2007).
16. K. Yagasaki, I. M. Merhasin, B. A. Malomed, T. Wagenknecht and A. R. Champneys, "Gap solitons in Bragg gratings with a harmonic superlattice," Europhys. Lett. **74**, 1006 (2006).
17. W.-H. Chen, Y.-J. He, and H.-Z. Wang, "Surface defect superlattice solitons," J. Opt. Soc. Am. B **24**, 2584 (2007).
18. R. Morandotti, D. Mandelik, Y. Silberberg, J. S. Aitchison, M. Sorel, D. N. Christodoulides, A. A. Sukhorukov, and Y. S. Kivshar, "Observation of discrete gap solitons in binary waveguide arrays," Opt. Lett. **29**, 2890 (2004).
19. M.I. Molina, I.L. Garanovich, A.A. Sukhorukov, and Y.S. Kivshar, "Discrete surface solitons in semi-infinite binary waveguide arrays", Opt. Lett. **31**, 2332 (2006)
20. K. Itoh, W. Watanabe, S. Nolte, and C. B. Schaffer, "Ultrafast Processes for Bulk Modification of Transparent Materials," MRS Bulletin **31**, 620 (2006).
21. A. Szameit, Y. V. Kartashov, V. A. Vysloukh, M. Heinrich, F. Dreisow, T. Pertsch, S. Nolte, A. Tünnermann, F. Lederer, and L. Torner, "Angular surface solitons in sectorial hexagonal arrays," Opt. Lett. **33**, 1542 (2008).
22. A. Szameit, Y. V. Kartashov, M. Heinrich, F. Dreisow, T. Pertsch, S. Nolte, A. Tünnermann, F. Lederer, V. A. Vysloukh, and L. Torner, "Observation of two-dimensional defect surface solitons," Opt. Lett. **34**, 797 (2009).
23. M. Heinrich, A. Szameit, F. Dreisow, R. Keil, S. Minardi, T. Pertsch, S. Nolte, and A. Tünnermann, "Observation of three-dimensional discrete-continuous X-waves in photonic lattices," Phys. Rev. Lett. **103**, 113903 (2009)

# References without titles

1. D. Christodoulides and R. Joseph, Opt. Lett. **13**, 794 (1988).
2. H. Eisenberg, Y. Silberberg, R. Morandotti, A. Boyd and J. Aitchison, Phys. Rev. Lett. **81**, 3383 (1998).
3. J. Fleischer, M. Segev, N. Efremidis and D. Christodoulides, Nature **422**, 147 (2003).
4. F. Lederer, G.I. Stegeman, D.N. Christodoulides, G. Assanto, M. Segev and Y. Silberberg, Phys. Rep. **463**, 1 (2008).
5. Y. V. Kartashov, V. A. Vysloukh, and L. Torner, Prog. Opt. **52**, 63 (2009).
6. X. Wang, A. Bezryadina, Z. Chen, K.G. Makris, D.N. Christodoulides, and G.I. Stegeman, Phys. Rev. Lett. **98**, 123903 (2007)
7. A. Szameit, Y.V. Kartashov, F. Dreisow, T. Pertsch, S. Nolte, A. Tünnermann, and L. Torner, Phys. Rev. Lett. **98**, 173903 (2007)
8. Y. S. Kivshar and N. Flytzanis, Phys. Rev. A **46**, 7972 (1992).
9. A. A. Sukhorukov and Y. S. Kivshar, Opt. Lett. **27**, 2112 (2002).
10. H. Ohno, E. E. Mendez, J. A. Brum, J. M. Hong, F. Agulló-Rueda, L. L. Chang, and L. Esaki, Phys. Rev. Lett. **64**, 2555 (1990).
11. J. Klos, Phys. Stat. Sol. B **242**, 1399 (2005).
12. M. Ghulinyan, C. J. Oton, Z. Gaburro, L. Pavesi, C. Toninelli, and D. S. Wiersma, Phys. Rev. Lett. **94**, 127401 (2005).
13. N. Malkova, I. Hromada, X. Wang, G. Bryant, and Z. Chen, Opt. Lett. **34**, 1633 (2009).
14. F. Dreisow, A. Szameit, M. Heinrich, T. Pertsch, S. Nolte, A. Tünnermann, and S. Longhi, Phys. Rev. Lett. **102**, 076802 (2009).
15. Y. J. He, W. H. Chen, H. Z. Wang, and B. A. Malomed, Opt. Lett. **32**, 1390 (2007).
16. K. Yagasaki, I. M. Merhasin, B. A. Malomed, T. Wagenknecht and A. R. Champneys, Europhys. Lett. **74**, 1006 (2006).
17. W.-H. Chen, Y.-J. He, and H.-Z. Wang, J. Opt. Soc. Am. B **24**, 2584 (2007).
18. R. Morandotti, D. Mandelik, Y. Silberberg, J. S. Aitchison, M. Sorel, D. N. Christodoulides, A. A. Sukhorukov, and Y. S. Kivshar, Opt. Lett. **29**, 2890 (2004).
19. M.I. Molina, I.L. Garanovich, A.A. Sukhorukov, and Y.S. Kivshar, Opt. Lett. 31, 2332 (2006)

20. K. Itoh, W. Watanabe, S. Nolte, and C. B. Schaffer, MRS Bulletin **31**, 620 (2006).
21. A. Szameit, Y. V. Kartashov, V. A. Vysloukh, M. Heinrich, F. Dreisow, T. Pertsch, S. Nolte, A. Tünnermann, F. Lederer, and L. Torner, Opt. Lett. **33**, 1542 (2008).
22. A. Szameit, Y. V. Kartashov, M. Heinrich, F. Dreisow, T. Pertsch, S. Nolte, A. Tünnermann, F. Lederer, V. A. Vysloukh, and L. Torner, Opt. Lett. **34**, 797 (2009).
23. M. Heinrich, A. Szameit, F. Dreisow, R. Keil, S. Minardi, T. Pertsch, S. Nolte, and A. Tünnermann, Phys. Rev. Lett. **103**, 113903 (2009).

# Figure captions

Figure 1.  $U$ versus $b$ for solitons residing in the (a) central sites of $D$- and $S$-lattices and (b) central ($M$), edge ($E$), and corner ($C$) sites of the $S$-lattice. Black, white and gray circles in (a) correspond to Figs. 2(g), 2(d) and 2(a), while in (b) they correspond to Figs. 2(d), 2(e), and 2(f).

Figure 2.  Solitons profiles in the $D$-lattice at (a) $b = 0.314$, (b) $b = 0.315$, and (c) $b = 0.311$. Soliton profiles in the $S$-lattice (d) $b = 0.327$, (e) $b = 0.325$, (f) $b = 0.322$, and (g)-(k) $b = 0.491$.

Figure 3.  Output patterns for (a) central, (b) edge, and (c) corner excitation in the $D$-lattice. First row: simulated linear patterns. Second row: observed linear pattern at input peak power of $200\,\mathrm{kW}$. Third and fourth row: observed nonlinear patterns at input peak powers of $1\,\mathrm{MW}$ and $2\,\mathrm{MW}$ respectively.

Figure 4.  Output patterns for the $S$-lattice. Arrangement corresponds to Fig. 3.

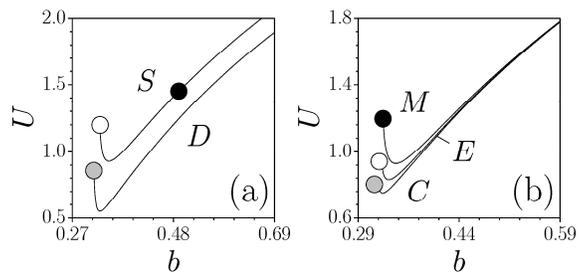

Figure 1.  $U$ versus $b$ for solitons residing in the (c) central sites of $D$- and $S$-lattices and (d) central $(M)$, edge $(E)$, and corner $(C)$ sites of the $S$-lattice. Black, white and gray circles in (c) correspond to Figs. 2(g), 2(d) and 2(a), while in (d) they correspond to Figs. 2(d), 2(e), and 2(f).

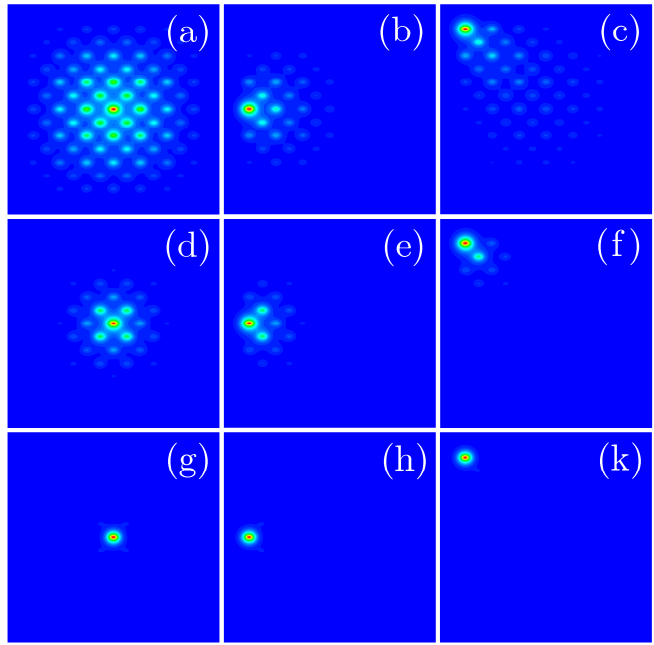

Figure 2. Solitons profiles in the $D$-lattice at (a) $b = 0.314$, (b) $b = 0.315$, and (c) $b = 0.311$. Soliton profiles in the $S$-lattice at (d) $b = 0.327$, (e) $b = 0.325$, (f) $b = 0.322$, and (g)-(k) $b = 0.491$.

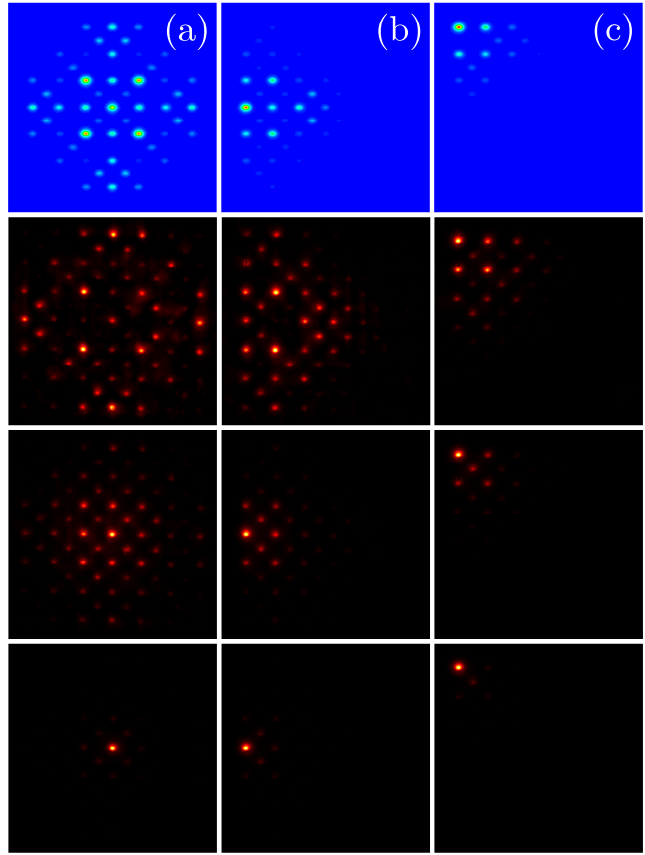

Figure 3. Output patterns for (a) central, (b) edge, and (c) corner excitation in the $D$-lattice. First row: simulated linear patterns. Second row: observed linear pattern at input peak power of 200 kW. Third and fourth row: observed nonlinear patterns at input peak powers of 1 MW and 2 MW respectively.

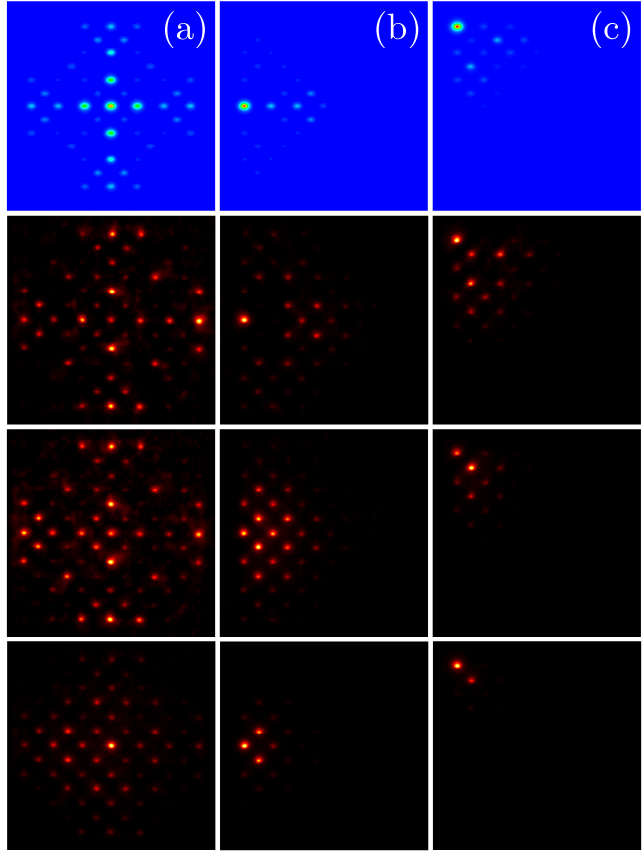

Figure 4. Output patterns for the $S$-lattice. Arrangement corresponds to Fig. 3.